\documentclass[twocolumn,showpacs,preprintnumbers,amsmath,amssymb]{revtex4}
\usepackage[dvips]{graphicx}
\usepackage{amssymb,amsfonts,amsmath}

\begin{document}

\title{Exact expression for the number of states in lattice models}
\author{Agata Fronczak}
\author{Piotr Fronczak}
\affiliation{Faculty of Physics, Warsaw University of Technology,Koszykowa 75, PL-00-662 Warsaw, Poland}

\date{\today}

\begin{abstract}
We derive a closed-form exact combinatorial expression for the number of
states in canonical systems with discrete energy levels. The expression
results from the exact low-temperature power series expansion of the
partition function. The approach provides interesting insights into basis of
statistical mechanics. In particular, it is shown that in some cases the
logarithm of the partition function may be considered the generating function
for the number of internal states of energy clusters, which characterize
system's microscopic configurations. Apart from elementary examples including
the Poisson, geometric and negative binomial probability distributions for
the energy, the framework is also validated against the one-dimensional Ising
model.
\end{abstract}


\maketitle

It has long been recognized that the number of states, $g(E)$, with a given
value of the energy, $E$, is a key quantity in equilibrium statistical
physics. In principle, all energy-related thermodynamic properties of a
classical canonical system can be calculated once $g(E)$ is known. In
particular, in the case of a system with discrete energy levels, the
canonical partition function is $Z(\beta)=\sum_{E}g(E)e^{-\beta E}$, where
the summation is over the allowed energy levels, and quantities such as the
Helmholtz free energy, $F(\beta)=-\ln Z(\beta)/\beta$, the ensemble-averaged
energy, $\langle E\rangle=-\partial \ln Z(\beta)/\partial\beta$, and the
specific heat (which is related to energy fluctuations, $\langle
E^2\rangle-\langle E\rangle^2=-\partial\langle E\rangle/\partial\beta$) can
be simply obtained.

Due to the central role of the energy distribution function, a variety of
theoretical and computational studies have addressed the problem of how to
obtain $g(E)$, see e.g. \cite{bookDombGreen, PRL1996Beale, PRL2001Wang,
PRL2004Micheletti, PRL2007Habeck, PRE2012Fronczak}. On the other hand,
relatively little was done to understand, how the state-space representation
of a considered system relates to its energy distribution, and how to recover
such an information from, for example, the partition function.

In this paper we derive a new and general combinatorial formula for the
density of states. The formula results from the exact low-temperature
power-series expansion of the canonical partition function. In our approach
$g(E)$ is expressed by the so-called Bell polynomials, which have a very
convenient combinatorial interpretation: they describe, how equal energy
portions can cluster together within the studied system. This, in turn, gives
insight into system's microscopic details.

In the low-temperature expansion of the partition function,
$Z(\beta)=\sum_{\Omega}e^{-\beta E(\Omega)}$, microstates, $\Omega$, are
counted in order of their importance as the temperature is increased from
zero \cite{inbookPlischke,inbookYeomans}. If the ground state of a system is
known and if the successive excitations from this state can be classified in
a simple way, one should, at least theoretically, be able to construct such a
series, i.e.
\begin{equation}\label{Z1a}
Z(x)=\sum_{N=0}^\infty g(\varepsilon N)x^N,
\end{equation}
where
\begin{equation}\label{Z1c}
x=e^{-\beta\varepsilon},
\end{equation}
and energy is considered to be discrete
\begin{equation}\label{Z1b}
E=\varepsilon N,
\end{equation}
with $\varepsilon$ representing a portion of energy and $N=0,1,2,\dots$ (with
the ground-state energy equal to zero).

In Eq.~(\ref{Z1a}), $g(\varepsilon N)$ represents the number of microscopic
configurations in which energy of the system is $\varepsilon N$.

From Eq.~(\ref{Z1a}) it is apparent that the exact distribution function for
the energy, $g(\varepsilon N)$, can be determined from the coefficients of
the partition function. As a rule, however, when the order of expansion is
increased complexity of contributing terms also increases rapidly. In this
context, a \emph{rule of thumb} states that the work involved in
\emph{direct} calculating the next term, $g(\varepsilon N)$, is the same as
that needed to calculate all the preceding terms,
$g(0),g(\varepsilon),\dots,g(\varepsilon N-\varepsilon)$. For this reason,
generation of lengthy low-temperature power-series for the partition function
is though to be a highly specialized art. That is why, even in the case of
the two-dimensional Ising model on a square lattice, for which the method was
primary advanced (see ch.~6 in~\cite{bookDombGreen}, the exact distribution
of energies was just obtained in 1996 \cite{PRL1996Beale,inbookPathria},
nearly a half of a century after Kaufman's exact solution for the partition
function \cite{PR1949Kaufman}.

The aim of this paper is to show that the effort needed to calculate the
coefficients of the low-temperature power-series expansion for the canonical
partition function is greatly reduced by using the combinatorial scheme
described below. The approach is very general and applies to any lattice
model. We also argue that the framework provides interesting insights into
basis of statistical mechanics. In particular, it is shown that in some cases
the logarithm of the partition function, $\ln Z(x)$, may be considered the
generating function for the number of internal states of energy clusters,
which characterize system's microscopic configurations. Apart from elementary
examples including the Poisson, geometric, and negative binomial probability
distributions for the energy, the framework is validated against the
one-dimensional Ising model, which is not at all a trivial example since its
ordering temperature is zero, and the series, Eq.~(\ref{Z1a}), is known to
diverge at $x=0$ as expected for a model with a zero temperature phase
transition \cite{inbookYeomans}.

\emph{Derivation of the main result.} To derive the main result of this
paper, which is the general combinatorial formula for the number of states,
we begin with the well-known relation between the canonical partition
function and the Helmholtz free energy, i.e.
\begin{equation}\label{Z2a}
Z(x)=e^{A(x)},
\end{equation}
where $x$ is the expansion variable of the low-temperature series,
Eq.~(\ref{Z1c}), and
\begin{equation}\label{Z2b}
A(x)=\ln Z(x)=\sum_{n=0}^\infty a_n\frac{x^n}{n!}=-\beta F(\beta),
\end{equation}
with $a_n=\partial^n A(x)/\partial x^n|_{x=0}$. Then, using the generating
function for Bell polynomials, $Y_{N}(\{a_n\})$, which are the polynomials
with a sequence of parameters $\{a_n\}=a_1,a_2,\dots a_N$, given by the
formal power series expansion \cite{inbookComtet}, Eq.~(\ref{Z2a}) can be
written as
\begin{eqnarray}\label{Z3a}
Z(x)=\exp\left[\sum_{n=0}^\infty a_n\frac{x^n}{n!}\right]=e^{a_0}\!\!\sum_{N=0}^\infty\frac{1}{N!}Y_N(\{a_n\})x^N.
\end{eqnarray}
In the last formula, the complete Bell polynomials, $Y_N(\{a_n\})$, are
defined as follows
\begin{equation}\label{Y1a}
Y_0(\{a_n\})=1,
\end{equation}
and for all $N\geq 1$
\begin{equation}\label{Y1}
Y_N(\{a_n\})=\sum_{k=1}^NB_{N,k}(\{a_n\}),
\end{equation}
where $B_{N,k}(\{a_n\})$ represent the so-called partial (or incomplete) Bell
polynomials, which can be calculated from the expression below
\begin{equation}\label{B1a}
B_{N,k}(\{a_n\})=N!\sum\prod_{n=1}^{N-k+1}\frac{1}{c_n!}\left(\frac{a_n}{n!}\right)^{c_n},
\end{equation}
where the summation takes place over all integers $c_n\geq 0$, such that
\begin{equation}\label{B1b}
\sum_{n=1}^{N-k+1}c_n=k\;\;\;\;\;\mbox{and}\;\;\;\;\;\sum_{n=1}^{N-k+1}nc_n=N.
\end{equation}
Finally comparing Eqs.~(\ref{Z1a}) and~(\ref{Z3a}), one gets the exact
expression for the number of states:
\begin{equation}\label{gE}
g(\varepsilon N)=\frac{e^{a_0}}{N!}Y_N(\{a_n\}).
\end{equation}

The basic difficulty with Eq.~(\ref{gE}) may arise from an unacquaintance
with Bell polynomials as given by Eqs.~(\ref{Y1}) and~(\ref{B1a}). For this
reason, we follow by explaining their meaning.

Suppose that $N$ distinguishable particles are partitioned into $k$ non-empty
and disjoint clusters of $n_i>0$ elements each, where $\sum_{i=1}^kn_i=N$.
There are exactly
\begin{equation}\label{Bell1}
{N\choose n_1,\dots,n_k}=N!\prod_{i=1}^k\frac{1}{n_i!}=
N!\!\!\prod_{n=1}^{N-k+1}\!\!\left(\!\frac{1}{n!}\!\right)^{c_n}
\end{equation}
of such partitions, where $c_n\geq 0$ stands for the number of clusters of
size $n$, with the largest cluster size being equal to $N-k+1$, and where
Eqs.~(\ref{B1b}) are satisfied. Suppose further that in such a composition
clusters of the same size are indistinguishable from one another, and each of
$c_n$ clusters of size $n$ can be in any one of $a_n\geq 0$ internal states.
Then the number of partitions becomes, cf. with Eq.~(\ref{Bell1}),
\begin{equation}\label{Bell2}
N!\prod_{n=1}^{N-k+1}\frac{1}{c_n!}\left(\frac{a_n}{n!}\right)^{c_n}.
\end{equation}
Summing Eq.~(\ref{Bell2}) over all integers $c_n\geq 0$ specified by
Eqs.~(\ref{B1b}), one gets the partial Bell polynomial, $B_{N,k}(\{a_n\})$,
which is defined by Eq.~(\ref{B1a}). If for all $n\geq 1$ the coefficients
$a_n\geq 0$, the polynomial describes the number of partitions of a set of
size $N$ with exactly $k$ subsets, where each coefficient, $a_n$, represents
the number of internal states of a cluster of size $n$. Finally, summing the
partial polynomials over $k$ one gets the complete Bell polynomial,
$Y_N(\{a_n\})$, the combinatorial meaning of which is obvious.

According to the explanations above the main result of this paper,
Eq.~(\ref{gE}), shows how equal energy portions, $\varepsilon$, are
distributed, and how they cluster together within the system whose structural
details are hidden in the coefficients $\{a_n\}$. If the coefficients are
non-negative, i.e. $\forall_{n\geq 1} a_n\geq 0$, they can be interpreted as
thermodynamic probabilities of energy clusters of a given size. Otherwise,
their meaning is not clear.

In what follows, the Poisson and the negative binomial distributions are
discussed as elementary mathematical examples of application of the
combinatorial approach. The advantages of the new framework with respect to
traditional methods of statistical mechanics are enhanced with reference to
the one-dimensional Ising model.


\emph{Poisson distribution.} As the first example, the Poisson distribution
for the energy of a canonical system is considered,
\begin{equation}\label{Poisson1}
P(E)=\frac{e^{-\langle E\rangle} \langle E\rangle^E}{E!},
\end{equation}
where, see Eq.~(\ref{Z1b}), $\varepsilon=1$ and $E=N=0,1,2,\dots$. In order
to bring Eq.~(\ref{Poisson1}) into the canonical form,
\begin{equation}\label{PE}
P(E)=g(E)\frac{e^{-\beta E}}{Z(\beta)},
\end{equation}
one has to assume that $\langle E\rangle=e^{-\beta}$. Then, the Poisson
distribution can be written as
\begin{equation}\label{Poisson2}
P(E)=\frac{1}{E!}\frac{e^{-\beta E}}{e^{e^{-\beta}}},
\end{equation}
where (compare~Eqs.~(\ref{PE}) and~(\ref{Poisson2}))
\begin{equation}\label{Poisson3}
g(E)=\frac{1}{E!}\;\;\;\;\;\mbox{and}\;\;\;\;\;\ln Z(\beta)=e^{-\beta}.
\end{equation}

To verify the combinatorial approach introduced in this paper, let us first
notice that in the case of the Poisson distribution $x=e^{-\beta}$,
Eq.~(\ref{Z1b}). Therefore, the logarithm of the partition function is
simply, cf.~Eqs.~(\ref{Z2b}) and~(\ref{Poisson3}),
\begin{equation}\label{Poisson4}
A(x)=x.
\end{equation}
The identity function for $A(x)$ provides a very significant sequence of the
coefficients $\{a_n\}$ for the energy distribution function, Eq.~(\ref{gE}),
i.e.
\begin{equation}\label{Poisson5}
a_0=0\;\;\;\;\;\mbox{and}\;\;\;\;\; \{a_n\}=1,0,0,\dots,0.
\end{equation}
The sequence states that in the case of a canonical system with the Poisson
distribution of energy, energy portions, $\varepsilon=1$, are independent
from each other. In some sense it means that such a system is structureless
(it may, for example, consists of noninteracting parts).

Finally, inserting the obtained coefficients into Eq.~(\ref{gE}) one can
easily get the energy distribution function, Eq.~(\ref{Poisson3}),
\begin{equation}\label{Poisson6}
g(N)=\frac{1}{N!}Y_N(1,0,0,\dots,0)=\frac{1}{N!},
\end{equation}
where $Y_N(1,0,0,\dots,0)=1$, because there is only one composition of a set
of size $N$ in which only clusters of size $n=1$ may exist ($a_1=1$ and
$a_n=0$ for all $n>1$).


\emph{Negative binomial distribution.} As the second example the negative
binomial distribution for the energy of a canonical system is studied,
\begin{equation}\label{NB1}
P(E)={E+r-1\choose E}(1-x)^rx^E,
\end{equation}
where $x\in(0,1)$ and $r>0$, while $\varepsilon=1$ and $E=N=0,1,2,\dots$, see
Eq.~(\ref{Z1b}). It is worth noting that among the discrete distributions,
the negative binomial distribution, $NB(r,x)$, Eq.~(\ref{NB1}), is considered
the discrete analogue of the Gamma distribution, $\Gamma(r,\beta)$, whose
probability density function is $P(E)=\beta^rE^{r-1}e^{-\beta E}/\Gamma(r)$,
where $r$ is the so-called shape parameter. The analogy is even more apparent
if one realizes that, in the case of $r=1$ the negative binomial distribution
becomes the geometric distribution, $NB(1,x)$, which is the discrete analogue
of the exponential distribution, $\Gamma(1,\beta)$.

Bringing the negative binomial distribution, Eq.~(\ref{NB1}), into the
canonical form, Eq.~(\ref{PE}), for $x=e^{-\beta}$ one gets:
\begin{equation}\label{NB3}
g(E)={E+r-1\choose E},
\end{equation}
and
\begin{equation}\label{NB4}
A(x)=-r\ln(1-x)=r\!\!\sum_{n=1}^{\infty}(n-1)!\frac{x^n}{n!}.
\end{equation}
The series expansion of the logarithm of the partition function, $A(x)$,
gives the following sequence of the coefficients $\{a_n\}$:
\begin{equation}\label{NB5}
a_0=0\;\;\;\;\;\mbox{and}\;\;\;\;\; \forall_{n\geq 1}\;a_n=r(n-1)!.
\end{equation}
The sequence provides a meaningful microscopic information about the
considered canonical system. First, it appears that the parameter $r$
describes the degeneracy of internal states of the energy-clusters, and the
lack of degeneracy, $r=1$, results in the geometric/exponential distribution.
Second, since $(n-1)!$ is the number of permutations of $n$ objects where the
first object is fixed, the only reasonable explanation behind
$a_n\propto(n-1)!$ is the linear ordering of energy portions within the
clusters. At first glance the idea of ordering of indistinguishable energy
portions may seem unacceptable, but it gets intelligibility if one realizes
that in some sense energy portions become distinguishable once they, for
example, become interaction energies between given pairs of particles.

To validate the main result of this paper, Eq.~(\ref{gE}), one has to insert
the coefficients given by  Eqs.~(\ref{NB5}) into the mentioned expression.
After some algebra one gets Eq.~(\ref{NB3}):
\begin{eqnarray}\label{NB7a}
g(N)&=&\frac{1}{N!}Y_{N}(\{r(n-1)!\})\\\label{NB7b}
&=&\frac{1}{N!}\sum_{k=1}^Nr^kB_{N,k}(0!,1!,2!,\dots)\\\label{NB7c}
&=&\frac{1}{N!}\sum_{k=1}^Nr^k{N\brack k}={N+r-1\choose N},
\end{eqnarray}
where basic combinatorial identities have been used, including
\cite{inbookComtet,bookStanley}: i. the generating function for the unsigned
Stirling numbers of the first kind, ${N\brack k}$, and ii. properties of Bell
polynomials, i.e. $B_{N,k}\left(\left\{ru^ka_n\right\}\right)=
r^ku^NB_{N,k}(\{a_n\})$ and $B_{N,k}(\{(n-1)!\})={N\brack k}$ \footnote{The
unsigned Stirling numbers of the first kind, ${N\brack k}$, count the number
of permutations of $N$ elements with $k$ disjoint cycles. Therefore,
$B_{N,k}(\{(n-1)!\})={N\brack k}$.}.


\emph{One-dimensional Ising model.} The Hamiltonian of the closed chain of
$V$ Ising spins, $\{s_i\}$, with nearest-neighbor interactions in the absence
of external magnetic field can be written as
\begin{equation}\label{IsingH}
H(\{s_i\})=-J\sum_{i=1}^Vs_is_{i+1},
\end{equation}
where $J>0$, $s_i=\pm 1$, and the periodic boundary condition is imposed by
assuming that $s_{V+1}=s_1$. In the thermodynamic limit, $V\gg 1$, the
partition function of the model is given by
\begin{equation}\label{IsingZ}
Z(\beta)=(e^{\beta J}+e^{-\beta J})^V.
\end{equation}

To use Eq.~(\ref{gE}) to calculate the number of states, $g(E)$,
characterizing the model, one must first rescale its energy, $E$, because in
the present form of the Hamiltonian, Eq.~(\ref{IsingH}), the ground state
energy is negative, $H(\{-1\})=H(\{+1\})=-JV$, whereas it should be at least
equal to zero, see Eqs.~(\ref{Z1a}),~(\ref{Z1b}) and~(\ref{Z1c}). The problem
can be solved by adding to Eq.~(\ref{IsingH}) a constant, $JV$, so that for
all the spin configurations, $\{s_i\}$, the modified Hamniltonian,
\begin{equation}\label{IsingH1}
H^*(\{s_i\})=-J\sum_{i=1}^Vs_is_{i+1}+JV,
\end{equation}
results in non-negative energy
\begin{equation}\label{IsingE1}
E^*=E+JV\geq 0.
\end{equation}
With this change, the partition function of the model becomes
\begin{equation}\label{IsingZ1}
Z^*(\beta)=e^{-\beta JV}Z(\beta)=(1+e^{-2\beta J})^V,
\end{equation}
but statistical properties of the chain of spins, $\{s_i\}$, remain the same,
as compared with the original Ising model. In particular, the probability
distributions for spin configurations, $P(\{s_i\})=\exp[-\beta
H(\{s_i\})]/Z(\beta)$ and $P^*(\{s_i\})=\exp[-\beta
H^*(\{s_i\})]/Z^*(\beta)$, coincide with one another. The same applies to
probability distributions for the energy, $P(E)=g(E)e^{-\beta E}/Z(\beta)$
and $P^*(E^*)=g^*(E^*)e^{-\beta E^*}/Z^*(\beta)$.

In view of the above considerations one can see that
\begin{equation}\label{IsinggE}
g(E)=g^*(E^*)=g^*(E+JV).
\end{equation}
Thus, to determine the number of states for the one-dimensional Ising model,
$g(E)$, one can calculate $g^*(E^*)$, and then change the variables $E^*$ and
$E$ according to Eq.~(\ref{IsingE1}).

In the following, for simplicity one assumes in Eq.~(\ref{IsingZ1}) that
$J=1/2$. Then, the logarithm of the partition function, $\ln Z^*(\beta)$ can
be written as, Eq.~(\ref{Z2b}),
\begin{equation}\label{IsingA}
A^*(x)=V\ln(1+x)=V\sum_{n=1}^\infty (-1)^{n-1}(n-1)!\frac{x^n}{n!},
\end{equation}
where $\varepsilon=1$ and $x=e^{-\beta}$. From the series expansion, it is
obvious that the coefficients $\{a_n\}$, i.e.
\begin{equation}\label{Isingan}
a_0=0\;\;\;\;\;\mbox{and}\;\;\;\;\;\forall_{n\geq 1} a_n=V(-1)^{n-1}(n-1)!,
\end{equation}
do not satisfy the condition of non-negativity, which (if satisfied) allows
to conclude on state space representation of the considered model.
Nevertheless, the knowledge of $\{a_n\}$ (regardless of their sign) enables a
direct validation of the combinatorial approach. By inserting the
coefficients into the expression for the energy distribution function,
Eq.~(\ref{gE}), after some algebra one gets, for $N=E^*$,
\begin{eqnarray}\label{Isingg0}
g^*(N)&=&\frac{1}{N!}Y_{N}(\{V(-1)^{n-1}(n-1)!\})\\\label{Isingg1}
&=&\frac{1}{N!}\sum_{k=1}^NV^k(-1)^{N-k}{N\brack k}={V\choose N},
\end{eqnarray}
where $(-1)^{N-k}{N\brack k}$ is the signed Stirling number of the first
kind. From the last expression one immediately gets
\begin{equation}\label{Isingg2}
g(E)={V\choose E+V/2},
\end{equation}
where $E$ is the energy of the original Ising model described by
Eqs.~(\ref{IsingH}) and~(\ref{IsingZ}).

The obtained result, Eqs.~(\ref{Isingg1}) and~(\ref{Isingg2}), is exactly the
expected one: the number of microscopic configurations of the closed chain of
$V$ Ising spins is equal to the number of ways to choose positions for $E^*$
pairs of equal spins, $\{+1,+1\}$ and $\{-1,-1\}$, from the available $V$
positions.

\emph{Summary.} Our approach differs crucially from previous works on the
number of states. We use the combinatorial mathematics of exponential
generating functions to obtain the exact low-temperature power-series
expansion of the canonical partition function. The expansion results in the
exact expression for the energy distribution in arbitrary lattice models.
Apart from elementary examples including the Poisson, geometric, and negative
binomial probability distributions for the energy, the framework is validated
against the one-dimensional Ising model. The work on the application of the
formalism to the Ising model on a square lattice is in progress and will be
reported in a future article to follow.

\emph{Acknowledgments.} The work has been supported from the National Science
Centre in Poland (grant no 2012/05/E/ST2/02300).


\end{document}